# Co-coupled synchronization of fractional-order unified chaotic systems

Ke-hui Sun [a, b*], Jian Ren [b], Shui-sheng Qiu [a]

[a] *School of Electronic and Information Engineering, South China University of Technology, Guangzhou 510641 China;*
[b] *School of Physics Science and Technology Central South University, Changsha 410083, China*

**Abstract**

Synchronization of fractional-order chaotic systems is a hot topic in the field of nonlinear study. The co-coupled synchronization between two fractional-order chaotic systems with different initial conditions is investigated in this paper. Based on Lyapunov stability principle and Gerschgorin theorem, the co-coupled synchronization theorem of fractional-order chaotic systems is deduced, and the range of coupling coefficients is confirmed for synchronization of fractional-order unified chaotic systems. By building up the synchronization simulation model on Simulink, the co-coupled synchronization between two fractional-order unified chaotic systems with different initial value is carried out, and the synchronization performances are analyzed, and the simulation results show that this synchronization method is effective.

PACS: 0545

## 1. Introduction

Fractional-order calculus theory has a 300-year-old history, but due to no practice application, it developed slowly. Until 1983 Mandelbort pointed out the fact that there are lots of fractional-order in nature and many fields of science and technology. Many systems are known to display fractional-order dynamics, such as viscoelastic systems [1], dielectric polarization [2], electrode-electrolyte polarization [3], electronmagnetic waves [4], quantitative finance [5], and quantum evolution of complex systems [6], so it is important to study the properties of fractional order systems.

Since the seminal paper by Pecora and Carroll in 1990 [7], chaos synchronization has attracted much attention [8] due to its theoretical challenge and its great potential applications in security communication [9-10]. Many synchronization approaches have been put forward, such as drive-response synchronization, feedback synchronization, co-couple synchronization, self-adaptive control synchronization, and active synchronization. Among those approaches, the co-couple synchronization has a bright future due to no requirement of decomposing the drive system and easy implementation.

Recently, synchronization of chaotic fractional differential systems starts to attract increasing attention and becomes hot topics [11-19]. Literature [17] discussed the synchronization of unified chaotic system based on states observer, and Literature [18] and [19] studied the synchronization of fractional-order Lü system and fractional-order Chen system respectively based on Fourier transform, but Fourier transform is just used to prove whether the fractional-order chaotic system can synchronize, but can not present the range of couple coefficients.

---

[*] Corresponding author.
*E-mail address:* kehui@csu.edu.cn (K. H. Sun)

In this article, we focus on the co-coupled synchronization of fractional order unified chaotic systems. The plan of the paper is as follows. In Section 2, we present the definition of fractional-order derivative and its approximation and the model of fractional-order unified system. In Section 3, we describe the synchronization principle of chaotic system based on co-coupled. In Section 4, chaos synchronization of the fractional-order unified system using the co-coupled approach is studied, and numerical simulations are used to show this process. Finally, we summarize the results and indicate future directions.

## 2. Fractional-order derivative and fractional-order unified chaotic system

### 2.1 Fractional-order derivative and its approximation

There are several definitions of fractional derivatives [20]. The best-known one is the Riemann-Liouvile definition, which is given by

$$\frac{d^{\alpha} f(t)}{dt^{\alpha}} = \frac{1}{\Gamma(n-\alpha)} \frac{d^n}{dt^n} \int_0^t \frac{f(\tau)}{(t-\tau)^{\alpha-n+1}} d\tau, \quad (1)$$

where $n$ is an integer such that $n-1 \leq \alpha < n$, $\Gamma(\cdot)$ is the *Gamma* function. The geometric and physical interpretation of the fractional derivatives was given in Ref.[21].

The Laplace transform of the Riemann-Liouville fractional derivative is

$$L\left\{\frac{d^{\alpha} f(t)}{dt^{\alpha}}\right\} = s^{\alpha} L\{f(t)\} - \sum_{k=0}^{n-1} s^k \left[\frac{d^{\alpha-1-k} f(t)}{dt^{\alpha-1-k}}\right]_{t=0}, \quad (2)$$

where, L means Laplace transform, and *s* is a complex variable. Upon considering the initial conditions to zero, this formula reduces to

$$L\left\{\frac{d^{\alpha} f(t)}{dt^{\alpha}}\right\} = s^{\alpha} L\{f(t)\}. \quad (3)$$

Thus, the fractional integral operator of order "$\alpha$" can be represented by the transfer function $H(s) = 1/s^{\alpha}$ in the frequency domain.

The standard definitions of fractional-order calculus do not allow direct implementation of the fractional operators in time-domain simulations. An efficient method to circumvent this problem is to approximate fractional operators by using standard integer-order operators. In Ref. [22], an effective algorithm is developed to approximate fractional-order transfer functions, which has been adopted in [22-24] and has sufficient accuracy for time-domain implementations. In Table 1 of Ref. [23], approximations for $1/s^{\alpha}$ with α from 0.1 to 0.9 in step 0.1 were given with errors of approximately 2 dB. We will use the $1/s^{0.95}$ approximation formula [24] in the following simulation examples.

$$\frac{1}{s^{0.95}} \approx \frac{1.2831s^2 + 18.6004s + 2.0833}{s^3 + 18.4738s^2 + 2.6574s + 0.003}. \quad (4)$$

### 2.2 Fractional-order unified chaotic system

In 2002, Lü J H *et al* put forward a new chaotic system [25], and it is described by

$$\begin{cases} \dot{x} = (25a+10)(y-x) \\ \dot{y} = (28-35a)x - xz + (29a-1)y \\ \dot{z} = xy - (8+a)z/3 \end{cases} \tag{5}$$

here, $a$ is the system parameter. When $a \in [0,1]$, there exists a chaotic attractor on *x-z* plane as shown in Fig.1(a).

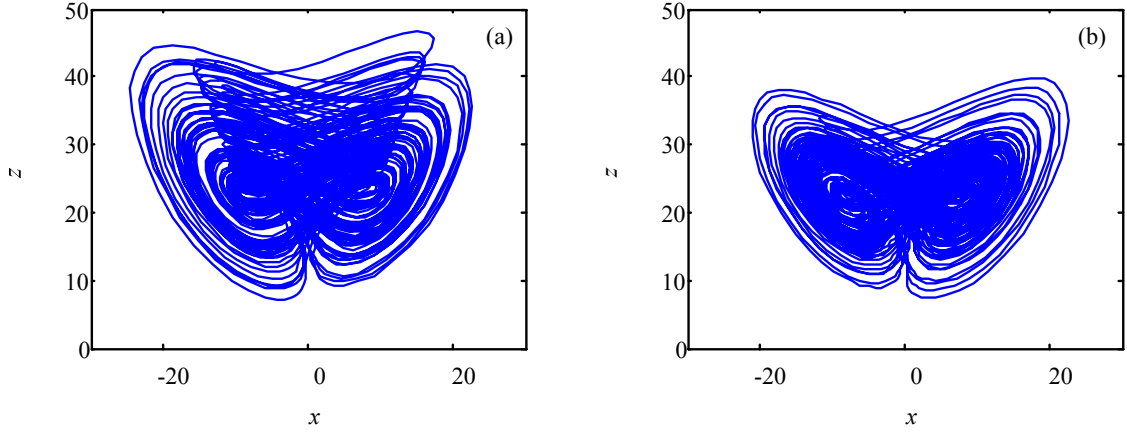

Fig.1 Chaotic attractor of unified system on *x-z* plane with *a*=1

(a) $\alpha=\beta=\gamma=1$  (b) $\alpha=\beta=\gamma=0.95$

Now, we consider the fractional-order unified system

$$\begin{cases} \dfrac{d^{\alpha} x}{dt^{\alpha}} = (25a+10)(y-x) \\ \dfrac{d^{\beta} y}{dt^{\beta}} = (28-35a)x - xz + (29a-1)y \\ \dfrac{d^{\gamma} z}{dt^{\gamma}} = xy - \dfrac{a+8}{3}z \end{cases} \tag{6}$$

where $0 < \alpha, \beta, \gamma < 1$, and system parameter $a \in [0,1]$. The lowest order existing chaos in this system is 2.76 [26]. A chaotic attractor in the system on *x-z* plane was shown in Fig.2(b) with $\alpha=\beta=\gamma=0.95$.

## 3. The principle of co-coupled synchronization of chaotic systems.

Consider a fractional-order chaotic system

$$\dfrac{d^{\alpha}\mathbf{X}}{dt} = A\mathbf{X} + f(\mathbf{X}), \tag{7}$$

where $\mathbf{X} \in R^n$ is the state vector, $A \in R^{n \times n}$ is the constant matrix of the linear part of the equation, and $f(\mathbf{X})$ is the nonlinear function of the system. Consider the co-coupled synchronization scheme of two fractional-order chaotic system

$$\dfrac{d^{\alpha}\mathbf{X}}{dt^{\alpha}} = A\mathbf{X} + f(\mathbf{X}) + D(\mathbf{X}' - \mathbf{X}), \tag{8}$$

$$\frac{d^\alpha \mathbf{X}'}{dt^\alpha} = A\mathbf{X}' + f(\mathbf{X}') + D(\mathbf{X} - \mathbf{X}'), \qquad (9)$$

here, $D = diag[d_1, d_2, \cdots, d_n]$ is the couple coefficients matrix of the synchronization system. Set $E = \mathbf{X} - \mathbf{X}'$, and $f(\mathbf{X}) - f(\mathbf{X}') = M_{\mathbf{X}, \mathbf{X}'}(\mathbf{X} - \mathbf{X}') = M_{\mathbf{X}, \mathbf{X}'}E$. Apparently, $M_{\mathbf{X}, \mathbf{X}'}$ is a bounded matrix, then one obtains the error system between (8) and (9) as follows,

$$\dot{E} = AE + M_{\mathbf{X}, \mathbf{X}'}E - 2DE = (A + M_{\mathbf{X}, \mathbf{X}'} - 2D)E. \qquad (10)$$

Obviously, $E=0$ is the equilibrium of system (10). This is to say that if an appropriate couple parameter matrix $D$ is chosen, the synchronization system will stable at the equilibrium point.

**Theorem 1** For the co-coupled driven system (8) and response system (9), let $D = diag[d_1, d_2, \cdots, d_n]$, $d_i \in R$, $i = 1, 2, \cdots, n$, $Q = (A + M_{\mathbf{X}, \mathbf{X}'})^T + (A + M_{\mathbf{X}, \mathbf{X}'})$, if $d_i > \frac{1}{4}(Q_{i,i} + \sum_{j=1, j\neq i}^{n} |Q_{i,j}|)$, then the error system (10) will converge, i.e. $\lim_{t \to +\infty} |E| = 0$, the global asymptotic synchronization between the driven system (8) and the response system (9) can be achieved.

**Proof** Constructing a Lyapunov function

$$V = E^T E, \qquad (11)$$

here $E = \mathbf{X} - \mathbf{X}'$. Apply differential to both side of this equation, we obtain

$$\dot{V} = \dot{E}^T E + E^T \dot{E} \qquad (12)$$

Then using Eq.(10), we have

$$\dot{V} = E^T[(A + M_{\mathbf{X}, \mathbf{X}'} - 2D)^T + (A + M_{\mathbf{X}, \mathbf{X}'} - 2D)]E = E^T H E, \qquad (13)$$

here, $H = (A + M_{\mathbf{X}, \mathbf{X}'} - 2D)^T + (A + M_{\mathbf{X}, \mathbf{X}'} - 2D)$. For unit matrix $U$ and real number matrix $\wedge = diag(\lambda_1, \lambda_2, \cdots, \lambda_n)$, we have $H = U^* \wedge U$, then

$$\dot{V} = E^T U^* \wedge U E = E'^T \wedge E', \qquad (13)$$

here, $E' = UE$. Let $Q = (A + M_{\mathbf{X}, \mathbf{X}'})^T + (A + M_{\mathbf{X}, \mathbf{X}'})$, then $H = Q - 4D$.

When $d_i > \frac{1}{4}(Q_{i,i} + \sum_{j=1, j\neq i}^{n} |Q_{i,j}|)$, then according to the well-known Gerschgorin's theorem in matrix theory, the eigenvalues of matrix $\wedge$ satisfied following formula

$$\lambda_i < 0 \qquad 1 \leq i \leq n. \qquad (14)$$

Then $\dot{V} < 0$, the error system (10) will converge, i.e. $\lim_{t \to +\infty} |E| = 0$. Then, the global asymptotic chaos synchronization between the driven system (8) and the response system (9) can be obtained.

## 4. Co-coupled synchronization of fractional-order unified chaotic system.

In this section, we build a co-coupled drive-response configuration with a drive system given by the fractional-order unified and with a response system given by its replica.

The drive system is given by

$$\begin{cases} \dfrac{d^\alpha x}{dt^\alpha} = (25a+10)(y-x) + d_1(x'-x) \\ \dfrac{d^\beta y}{dt^\beta} = (28-35a)x - xz + (29a-1)y + d_2(y'-y) \\ \dfrac{d^\gamma z}{dt^\gamma} = xy - \dfrac{a+8}{3}z + d_3(z'-z) \end{cases} \quad (15)$$

and the response one is defined as

$$\begin{cases} \dfrac{d^\alpha x'}{dt^\alpha} = (25a+10)(y'-x') + d_1(x-x') \\ \dfrac{d^\beta y'}{dt^\beta} = (28-35a)x' - x'z' + (29a-1)y' + d_2(y-y') \\ \dfrac{d^\gamma z'}{dt^\gamma} = x'y' - \dfrac{a+8}{3}z' + d_3(z-z') \end{cases} \quad (16)$$

here, $d_1$, $d_2$, $d_3$ are the couple coefficients. Set $e_x = x-x'$, $e_y = y-y'$, $e_z = z-z'$, then we obtains the error system between (15) and (16) as follows

$$\begin{cases} \dfrac{d^\alpha e_x}{dt^\alpha} = (25a+10)e_y - (25a+10-2d_1)e_x \\ \dfrac{d^\beta e_y}{dt^\beta} = (28-35a)e_x - z'e_x - xe_z + (29a-1-2d_2)e_y \\ \dfrac{d^\gamma e_z}{dt^\gamma} = y'e_x + xe_y - [(8+a)/3 + 2d_3]e_z \end{cases} \quad (17)$$

Let $\alpha = \beta = \gamma = 0.95$, then we have

$$\dot{E} = AE + M_{\mathbf{X},\,\mathbf{X'}}E - 2DE = (A + M_{\mathbf{X},\,\mathbf{X'}} - 2D)E \quad (18)$$

where

$$A = \begin{bmatrix} -(25a+10) & 25a+10 & 0 \\ 28-35a & 29a-1 & 0 \\ 0 & 0 & -\dfrac{8+a}{3} \end{bmatrix},\ D = \begin{bmatrix} d_1 & 0 & 0 \\ 0 & d_2 & 0 \\ 0 & 0 & d_3 \end{bmatrix},\ E = \begin{bmatrix} x-x' \\ y-y' \\ z-z' \end{bmatrix},\ f(\mathbf{X}) = \begin{bmatrix} 0 \\ -xz \\ xy \end{bmatrix},$$

$$f(\mathbf{X'}) = \begin{bmatrix} 0 \\ -x'z' \\ x'y' \end{bmatrix}, \quad f(\mathbf{X}) - f(\mathbf{X'}) = \begin{bmatrix} 0 \\ -xz + x'z' \\ xy - x'y' \end{bmatrix} = \begin{bmatrix} 0 & 0 & 0 \\ -z' & 0 & -x \\ y' & x & 0 \end{bmatrix} \begin{bmatrix} x - x' \\ y - y' \\ z - z' \end{bmatrix}, \quad M_{\mathbf{X}, \mathbf{X'}} = \begin{bmatrix} 0 & 0 & 0 \\ -z' & 0 & -x \\ y' & x & 0 \end{bmatrix}$$

So

$$Q = (A + M_{\mathbf{X}, \mathbf{X'}})^T + (A + M_{\mathbf{X}, \mathbf{X'}}) = \begin{bmatrix} -50a - 20 & -10a + 38 - z' & y' \\ -10a + 38 - z' & 58a - 2 & 0 \\ y' & 0 & -\dfrac{16 + 2a}{3} \end{bmatrix}. \quad (19)$$

According to Theorem 1, when $d_i > \dfrac{1}{4}(Q_{i,i} + \sum\limits_{j=1, j \ne i}^{n} |Q_{i,j}|)$, i.e. $d_1 > \dfrac{1}{4}((-50a - 20) + |-10a + 38 - z'| + |y'|)$,

$d_2 > \dfrac{1}{4}((58a - 2) + |-10a + 38 - z'|)$, $d_3 > \dfrac{1}{4}(-\dfrac{16 + 2a}{3} + |y'|)$, the synchronization can be achieved. For

the fractional-order unified chaotic system, we have $|y'| < 40$, $|z'| < 50$, then we can obtain that if

$d_1 > 27 - 15a$, $d_2 > 12a + 22$, $d_3 > (52 - a)/6$, then the synchronization between systems (15) and (16)

is implemented. For example, if let $a=0.8$, then $d_1 > 15$, $d_2 > 31.6$, $d_3 > 8.5$. According to the principle

of co-couple synchronization, a simulation model is built up on the platform of Simulink, which is shown
in Fig.2.

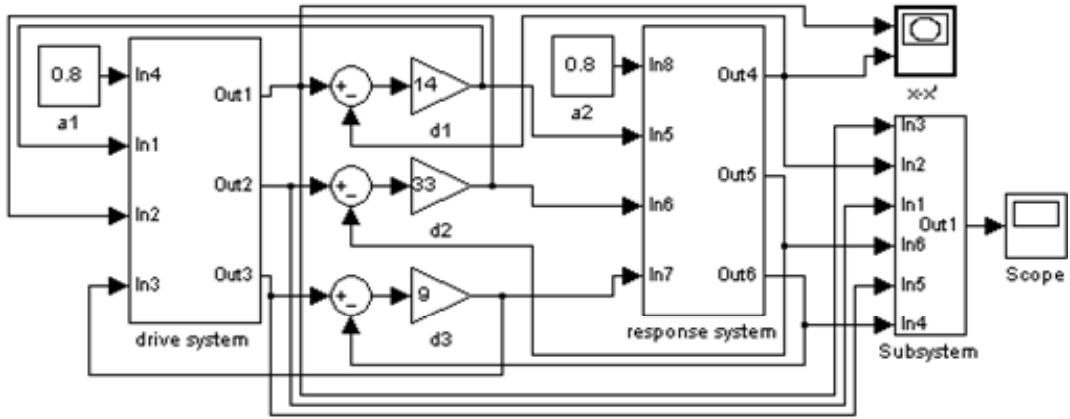

Fig.2 Co-couple synchronization simulation model

When $a = 0.8$, $\alpha = \beta = \gamma = 0.95$, system（16）and system（17）are chaotic. Using the synchronization simulation model, Synchronization simulation is complemented. Here, the initial values of driven and response system are chosen as [$x(0), y(0), z(0)$]=[1,1,1], [$x'(0), y'(0), z'(0)$]=[15,15,15] respectively, and set $d_1 = 16, d_2 = 33, d_3 = 9$, and $|E| = \sqrt{(x' - x)^2 + (y' - y)^2 + (z' - z)^2}$. Simulation time is 20 seconds. The simulation results is present in Fig.3. Obviously, two fractional-order unified chaotic systems with different initial values can be synchronized quickly.

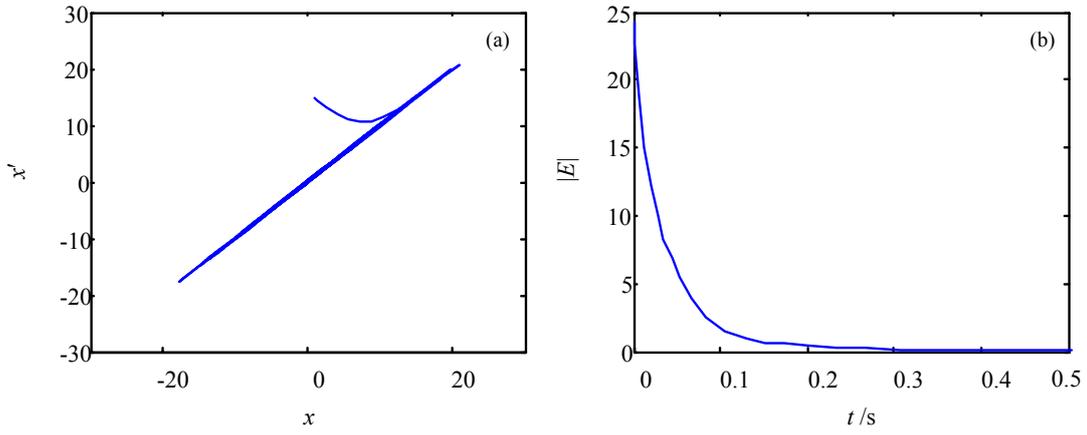

Fig.3 Synchronization simulation diagram of fractional-order unified chaotic system

(a) Synchronization phase diagram with $x$-$x'$  (b) Synchronization error curve

For the purpose of comparison, we also plot the curves of synchronization error of the integer-order unified chaotic systems in Fig.4 (Except $\alpha = \beta = \gamma = 1$, all the experiment conditions are the same with Fig.3). Comparing Fig.3 with Fig.4, it is found that the synchronization rate of the fractional-order chaotic oscillators is a little bit slower than its integer-order counterpart.

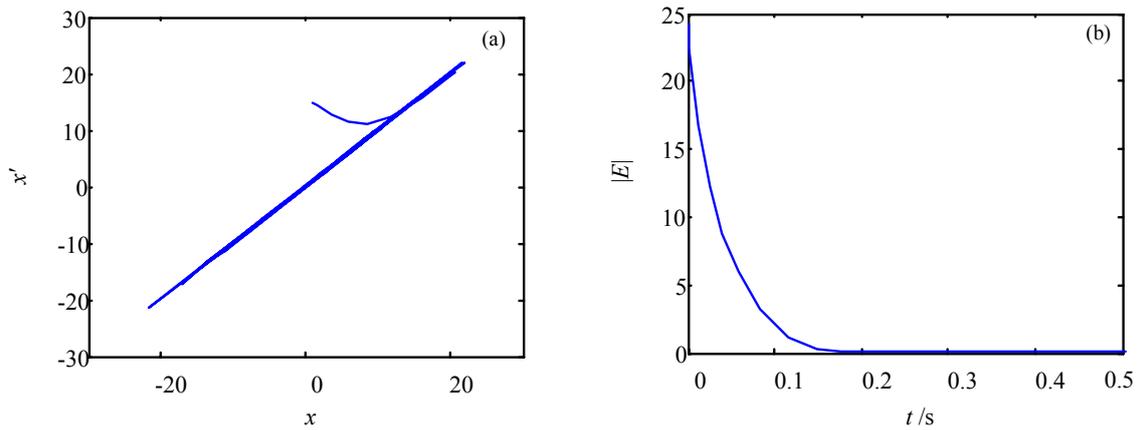

Fig.4 Synchronization simulation diagram of integer-order unified chaotic system

(a) Synchronization phase diagram with $x$-$x'$  (b) Synchronization error curve

To analyze the synchronization performance of this synchronization approach, simulation experiments are carried out with different co-couple strengths, and the results are present in Fig.5. It is noticed that the upbuilding time required for achieving synchronization between the drive system and the response one sensitively depend on the coupling strength. Generally, the synchronization upbuilding time increases with the co-couple strength increasing. It is useful for us to apply this synchronization approach to secure communication.

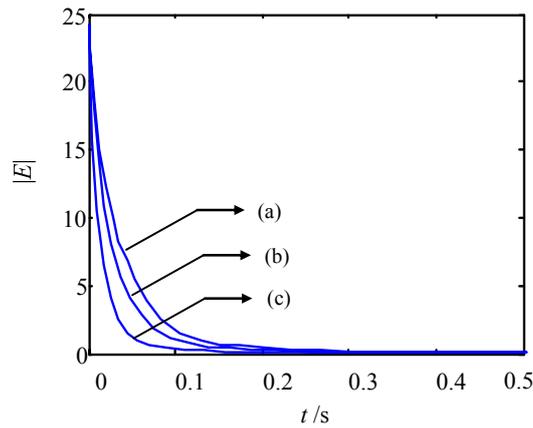

Fig. 5 Synchronization error curve with different co-couple strength
(a) $d_1=16$, $d_2=33$, $d_3=9$   (b) $d_1=18$, $d_2=37$, $d_3=13$   (b) $d_1=30$, $d_2=49$, $d_3=25$

## 5. Conclusions

Based on the theorem of Lyapunov stability and Gerschgorin theorem, we have studied the co-coupled synchronization between two fractional-order unified chaotic systems with different initial values. The sufficient conditions of synchronization is deduced, especially the range of coupling coefficient is obtained. We find that two fractional-order chaotic oscillators can be brought to an exact synchronization with appropriate coupling strength. Chaotic synchronization in fractional-order systems is intricate. Future work includes the application of fractional-order unified chaotic system in the secure communication.

## Acknowledgments

This work was supported by the National Nature Science Foundation of People's Republic of China (Grant No. 60672041), and the National Science Foundation for Post-doctoral Scientists of People's Republic of China (Grant No. 20070420774).